# The role of vector potential coupling in hot electron cooling power in bilayer graphene at low temperature

S. S. Kubakaddi

Department of Physics, K. L. E. Technological University, Hubballi-580031, Karnataka, India
Email: sskubakaddi@gmail.com
( June 24, 2017)

We have studied, in bilayer graphene (BLG), the hot electron cooling power $F_{VP}(T, n_s)$ due to acoustic phonons via vector potential (VP) coupling. It is calculated as a function of electron concentration $n_s$ and temperature $T$ and compared with $F_{DP}(T, n_s)$ that due to deformation potential coupling. For the $n_s$ around $1 \times 10^{12}$ cm$^{-2}$, $F_{VP}(T, n_s)$ is much smaller than $F_{DP}(T, n_s)$. With increase of $n_s$, $F_{DP}(T, n_s)$ decreases faster than $F_{VP}(T, n_s)$ does. A cross over is predicted and dominant contribution of $F_{VP}(T, n_s)$ can be observed at large $n_s$. In the Bloch- Grüneisen (BG) regime $F_{VP}(T, n_s) \sim n_s^{-1/2}$ and $F_{DP}(T, n_s) \sim n_s^{-3/2}$. Both $F_{VP}(T, n_s)$ and $F_{DP}(T, n_s)$ have the same $T$ dependence with $T^4$ power law in BG regime. Behaviour of $F_{DP}(T, n_s) \sim n_s^{-3/2}$ and $T^4$ is in agreement with the experimental results at moderate $n_s$. Besides, in BG regime, we have predicted, for both the VP and DP coupling, a relation between $F(T, n_s)$ and the acoustic phonon limited mobility $\mu_p$, opening a new door to determine $\mu_p$ from the measurements of $F(T, n_s)$

## 1. Introduction

Hot electron dynamics in graphene has been intensively investigated both theoretically and experimentally [1-14]. It is of considerable importance due to its applications in high speed devices, bolometry and calorimetry. The electrons in graphene get heated by the electric field or by the incident radiation and establish their 'electron temperature $T$' greater than the lattice temperature $T_L$. In the steady state, these hot electrons lose their energy through electron-phonon interaction as the cooling channel with the emission of acoustic (optical) phonons at low (high >200-300 K) temperature. In graphene, due to the weak electron-acoustic phonon (el-ap) coupling and large optical phonon energy (~200 meV), the hot electrons remain thermally decoupled from the lattice over a wide range of low temperature. This suppressed thermal link between electrons and phonons make graphene as a better potential candidate in its application for high speed devices and bolometry. In view of this, a rigorous and quantitative understanding of the electron-phonon interaction is being made at low temperatures in monolayer graphene (MLG) (for eg see Refs.[8,9,13]).

In graphene, the electron-acoustic phonon interaction is via the deformation potential (DP) coupling (which is also called scalar coupling). In the Bloch- Grüneisen (BG) regime the temperature dependence of electron cooling power $P$, due to the unscreened DP coupling, is predicted and observed to be given by the power law $P \sim T^4$ ($T^3$) in clean ( dirty) limit [1,3,4-10,13]. In the samples with disorder, the cooling is due to disorder assisted el-ap 'supercollision' [5-7,10]. Quantitative agreement has been obtained between the measured and calculated values. In MLG, there exists some theoretical studies of resistivity [15] and hot electron heat flux transfer [10] due to the vector potential (VP) coupling of el-ap interaction, which predict qualitatively the same $T$ dependence as that of the unscreened DP coupling. In suspended MLG samples, additional electron cooling by the emission of flexural phonons is also discussed in Ref. [13], which is ignored in the samples grown on the substrate.

There are limitations on the range of applicability of MLG in electronic devices because of its zero-energy gap.

Bilayer graphene (BLG) is potential because it shows a tuneable energy gap, and a parabolic dispersion relation with finite effective mass [16, 17]. The existence of variable energy gap makes BLG the most promising material for fabrication of graphene electronic devices and radiation sensors. There are relatively few hot electron cooling power studies in BLG. Electron cooling by emission of acoustic phonons due to DP coupling [4,18], surface phonons [19] and optical phonons [4] is investigated theoretically. The theoretical predictions [18], in the BG regime, give the temperature dependence $\sim T^4$ and electron concentration dependence $\sim n_s^{-3/2}$ contrary to the $n_s^{-1/2}$ dependence in MLG [1].

Nicholas and co-workers [20] have experimentally investigated the energy loss rate $P$ of hot electrons in epitaxial bilayer graphene. The observed $P$ is found to follow the predicted BG power law behavior of $T^4$ (temperature up to ~100 K.) and the electron concentration dependence $\sim n_s^{-3/2}$. They have also observed, in contrast, the $n_s^{-1/2}$ dependence in MLG leading to a cross over in the energy loss rate as a function of electron density between these two systems. These observations are shown to be in good agreement with the theoretical predictions due to electron-acoustic phonon interactions with the unscreened DP coupling constant of 22 eV. Contributions due to supercollisions and vector potential coupling are not noticed in their observations in BLG. The dependence of $P \sim n_s^{-3/2}$ contrary to the $n_s^{-1/2}$ dependence in MLG [1] is the important difference which is exploited to distinguish Dirac phase of electrons from 2DEG with quadratic dispersion [20].

Electron-phonon heat transfer rate is also investigated experimentally and theoretically in a suspended BLG, near Dirac point, for $T = 10-1000$ K by Laitinen et al. [21]. These authors have obtained a good agreement with their theoretical estimates by considering the heat conduction flow governed by Wiedemann-Franz law in the low temperature region and zone edge and zone center optical phonons at higher temperatures. Their theoretical estimates of heat flow due to the DP coupling of acoustic phonons are found to be well below the observed values. Their study rules out 'supercollision' processes in BLG.

In BLG, the contribution and importance of VP coupling to the temperature dependence of resistivity is investigated by Ochao et al. [15]. In their treatment the VP coupling contribution is shown to be much higher than that of the screened DP coupling. In the present work, we address, in bilayer graphene, the contribution of hot electron cooling power due to electron-acoustic phonon interaction via VP coupling. We investigate its concentration $n_s$ and temperature $T$ dependence and compare it with those due to DP coupling. The power laws are predicted in BG regime with regard to $n_s$ and $T$ dependence. Our calculations emphasize more on $n_s$ dependence of $P$. The circumstances under which contribution due to the VP coupling can be important are explored. Also, we bring out a simple but important relation between the hot electron cooling power and the acoustic phonon limited mobility in Bloch-Gruneisen regime, which will enable to extract the latter from the measurements of the former.

## 2. Theory

Basic equation for hot electron cooling power is given by $P = (1/N_e)\sum_{\mathbf{q}} \hbar\omega_{\mathbf{q}}(dN_{\mathbf{q}}/dt)$, [1] where $N_e$ is the total number of electrons and $(dN_{\mathbf{q}}/dt)$ is the rate of change of phonon distribution function equation $N_{\mathbf{q}}$ due to el-ap phonon interaction. The electron states and energy dispersion $E_{\mathbf{k}}$ are given in Ref. [16]. In the following we give the equation for cooling power, which is common to both the VP and DP coupling. Finally, the substitutions are made for the respective coupling constants and the functions arising due to spinors.

The Matrix elements for the el-ap interaction, causing the scattering between the initial electron state $\mathbf{k}$ and final state $\mathbf{k'}$, are given by $|M_{DP}(q)|^2 = D^2 \xi(q) g_{DP}(\theta_{\mathbf{k},\mathbf{k'}})$ and $|M_{VP}(q)|^2 = D_1^2 \xi(q) g_{VP}(\mathbf{k},\mathbf{k'})$, respectively, for DP and VP coupling [15]. In these equations, $D$ is the deformation potential coupling constant, $\xi(q) = (\hbar\omega_{\mathbf{q}})/(2\rho A v_s^2)$, $D_1 = (\hbar^2\beta/4ma)$, $g_{DP}(\theta_{\mathbf{k},\mathbf{k'}}) = (1+\cos 2\theta_{\mathbf{k},\mathbf{k'}})/2$, $g_{VP}(\mathbf{k},\mathbf{k'}) = (k^2 + k'^2 + 2kk'\cos\theta_{\mathbf{k},\mathbf{k'}})/2$, $\rho$ is the areal mass density of BLG, $A$ is the surface area, $\omega_{\mathbf{q}}$ is the frequency of acoustic phonons of wave vector $\mathbf{q}$, $v_s$ is the velocity of the acoustic phonons (subscript $s=l$ for longitudinal and $s=t$ for transverse), $v_f$ is the Fermi velocity of Dirac electrons, $\beta$ ($\approx$ 2-3) is the vector potential gauge parameter, $m$ is the effective mass of the electron and $a$ is the distance between the carbon atoms.

$P$ is conveniently expressed as $P = F(T, n_s) - F(T_L, n_s)$, where $F(T, n_s)$ is found to be

$$F_i(T,n_s) = \frac{m^{3/2} g D_i^2}{2^{5/2} \pi^2 \rho n_s \hbar^5 v_s^3} \int_0^\infty d(\hbar\omega_q)(\hbar\omega_q)^2 N_{\mathbf{q}}(T)$$
$$\times \int_{\gamma(q)}^\infty dE_{\mathbf{k}} \frac{g_i(q,k)}{\sqrt{(E_{\mathbf{k}} - \gamma(q))}} [f(E_{\mathbf{k}} + \hbar\omega_{\mathbf{q}}) - f(E_{\mathbf{k}})]. \quad (1)$$

Here, the subscript $i$ = DP and VP, $g = 4$ is the product of spin ($g_s=2$) and valley ($g_v=2$) degeneracy, $D_i = D$ and $D_1$, respectively, for DP and VP coupling, $n_s$ is the 2D electron concentration, $f(E_{\mathbf{k}})$ is the electron distribution function at temperature $T$ and $\gamma(q) = (\hbar\omega_q - E_q)^2/4E_q$, The $g_i(q,k) = g_{DP}(q,k)$ and $g_{VP}(q,k)$, respectively, for DP and VP coupling. They are given by $g_{DP}(q,k) = [1+(\hbar\omega_{\mathbf{q}} - E_{\mathbf{k}})/2E_{\mathbf{k}}]^2/[1+(\hbar\omega_{\mathbf{q}}/E_{\mathbf{k}})]$ and $g_{VP}(q,k) = (4mE_{\mathbf{k}}/\hbar^2)[1+(\hbar\omega_{\mathbf{q}}/2E_{\mathbf{k}}) - (E_q/4E_{\mathbf{k}})]$. These expressions for $g_{DP}(q,k)$ and $g_{VP}(q,k)$ are taken in inelastic regime. Normally, these equations are taken in quasi-elastic approximation in which $\hbar\omega_{\mathbf{q}}$ is ignored with respect to the electron energy $E_{\mathbf{k}}$.

In the above equation we note that DP coupling is only with LA phonons and we use $v_s = v_l$. In case of the VP coupling, both LA and TA(in-plane) phonons are involved and $v_s = v_l$ and $v_t$ are taken. Moreover, we note that $P = F_i(T, n_s)$ for $T >> T_L$ and for $T_L = 0$. Although, in the literature [1,4,22], conventionally $F(T)$ is used to denote the electron cooling power at temperature $T$, we prefer to use $F(T, n_s)$, as our discussion will be with emphasis on $n_s$ dependence.

In the BG regime ($\hbar\omega_{\mathbf{q}} << E_f$, the Fermi energy), we obtain the following power laws

$$F_{DP}(T,n_s) = F_{DP0} \frac{T^4}{n_s^{3/2}}, \quad (2)$$

with

$$F_{DP0} = \frac{gD^2 m^2 k_B^4 3!\zeta(4)}{4\rho\pi^{5/2} \hbar^6 v_s^3}, \quad (2a)$$

for DP coupling, and

$$F_{VP}(T,n_s) = F_{VP0} \frac{T^4}{n_s^{1/2}}, \quad (3)$$

with

$$F_{VP0} = \frac{gD_1^2 m^2 k_B^4 3!\zeta(4)}{2\rho\pi^{3/2} \hbar^6 v_s^3}, \quad (3a)$$

for VP coupling. Here $\zeta(n)$ is Riemann zeta function.

## 3. Results and discussion

We discuss the contribution of both t DP and VP coupling to the electron cooling power $F(T)$, independently and their sum, as a function of electron concentration $n_s$ and temperature $T$. In order to evaluate the cooling power (Eq.(1)) numerically, we chose the parameter values typically used in the literature: $\rho_s = 15.2 \times 10^{-8}$ gm/cm$^2$, $D = 20$ eV [[1,8-10], $\beta = 2.5$ [15], $m = 0.035$ $m_0$ [4], $a = 1.4$ Å, $v_f = 1 \times 10^8$ cm/s, $v_l = 2 \times 10^6$ cm/s and , $v_t = 1.4 \times 10^6$ cm/s [15]. We point out that, in the literature, the most commonly used values of $D$ are closer to 20 eV, although the quoted range is 3-30 eV. The calculations are presented for zero lattice temperature and without screening of DP coupling. The el-ap interaction via VP coupling is not screened [15]. The BG regime is defined by the temperature region $< T_{BG} = 2 \hbar v_s k_f/k_B$, where $k_f$ is the Fermi wave vector. $T_{BG} = 54.153\sqrt{n_s}$ K and $37.9\sqrt{n_s}$, respectively, for LA and TA phonons, with $n_s$ taken in units of $10^{12}$ cm$^{-2}$. The power law equations (2) and (3) are valid strictly for $T << T_{BG}$.

### 3.1. Electron concentration dependence of $F(T,n_s)$

In Fig. 1a, $F(T,n_s)$ is shown as a function of $n_s$ ( 0.5-5x10$^{12}$ cm$^{-2}$) at $T = 1$ K. Both $F_{DP}(T,n_s)$ and $F_{VP}(T, n_s)$ decrease with increasing $n_s$. The $F_{DP}(T,n_s)$ decreases more rapidly than $F_{VP}(T, n_s)$ does. For $n_s = 0.5 \times 10^{12}$ cm$^{-2}$ $F_{DP}(T,n_s)$ is found to be nearly thirty times the $F_{VP}(T, n_s)$. For $n_s$ up to about 3.0x10$^{12}$ cm$^{-2}$, the contribution of $F_{VP}(T, n_s)$ is almost negligible. In the larger $n_s$ region, the total $F_T(T,n_s) = F_{DP}(T,n_s) + F_{VP}(T, n_s)$ decreases less rapidly than $F_{DP}(T,n_s)$ alone does. This is attributed to the weak contribution due to $F_{VP}(T, n_s)$ adding up at larger $n_s$. We have also shown the respective curves from the power laws (Eqs.(2) and (3)) of BG regime. At such low temperature $T= 1$ K, BG regime curves are coinciding with their respective curves drawn from Eq.(1). The behaviour of $F_{DPBG}(T,n_s) \sim n_s^{-3/2}$ and $F_{VPBG}(T, n_s) \sim n_s^{-1/2}$ indicate the stronger $n_s$ dependence due to DP coupling in BLG. The experimental observations of Nicholas et al [20] show $n_s^{-3/2}$ dependence in the range $n_s$ =1-3x$10^{12}$ cm$^{-2}$, in perfect agreement with the DP coupling contribution. As mentioned above, in this region of $n_s$ the contribution from the VP coupling is relatively insignificant. It is to be noted that in the $n_s$ region where $F_{VP}(T, n_s)$ is becoming important, the power of $n_s$ decreases from -3/2 to lower value. This can be seen from the behaviour of the total $F_T(T,n_s)$ in the larger $n_s$ region.

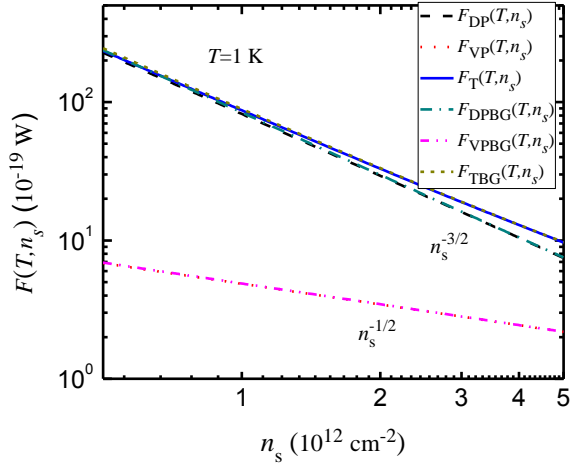
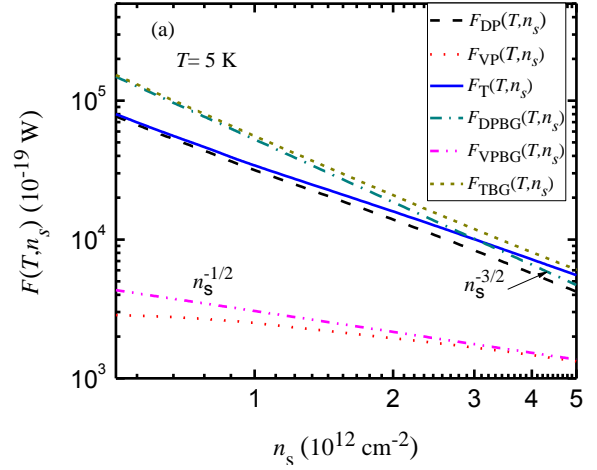

**Figure 1:** $F(T,n_s)$ vs $n_s$ for $T=1$ K due to DP and VP coupling and their total $F_T(T,n_s)= F_{DP}(T,n_s)+ F_{VP}(T,n_s)$. BG regime curves are also shown with their total $F_{TBG}(T,n_s)= F_{DPBG}(T,n_s)+ F_{VPBG}(T,n_s)$. These BG law curves are coinciding with the respective full equation curves.

In Fig. 2a, $F(T,n_s)$ vs $n_s$ are shown at $T= 5$ K. The curves are shown for $F_{DP}(T,n_s)$, $F_{VP}(T, n_s)$, their total $F_T(T,n_s)$ and the respective curves of the BG regime power law. There is difference in the curves from Eq.(1) and the respective curves from the BG regime power law equations, as power laws become less valid for larger $T$. This difference deceases as $n_s$ increases, as expected, due to the more validity of power law for the larger $n_s$. Merging of the respective curves for VP coupling occurs, relatively, at lower $n_s$ (around 3.0x$10^{12}$ cm$^{-2}$) whereas DP coupling curves seem to merge beyond 5.0x$10^{12}$ cm$^{-2}$.

Due to the difference in the $n_s$ dependence of DP and VP coupling contributions, as observed in Fig. 1 and Fig. 2a, we expect the cross over between $F_{DP}(T,n_s)$ and $F_{VP}(T, n_s)$ when they are plotted over a large range of $n_s$. In order to see the cross over, they are shown in Fig. 2 over the range of $n_s = 0.5$-50x$10^{12}$ cm$^{-2}$ at $T= 5$ K. We find this cross over taking place for about $n_s$=2.0x$10^{13}$ cm$^{-2}$. Interestingly, our calculations show, even for $T= 50$ K, the cross over takes place at around the same $n_s$. For the $n_s$ well above this cross over region, observations are expected to be dominated by $F_{VP}(T, n_s)$ and follow $n_s^{-1/2}$ behaviour. We suggest, in order to see the VP coupling contribution to the hot electron cooling power, measurements need to be made over the large range of $n_s$ covering the cross over region.

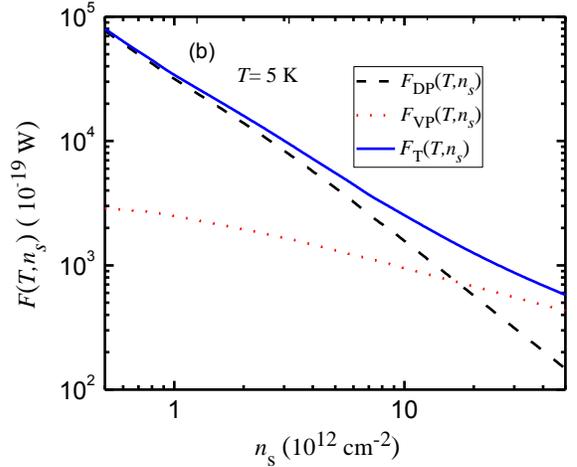

**Figure 2:** $F(T,n_s)$ vs $n_s$ for $T=5$ K due to DP and VP coupling and their total $F_T(T,n_s)= F_{DP}(T,n_s)+ F_{VP}(T,n_s)$. (a) for $n_s$=0.5-5.0x$10^{12}$ cm$^{-2}$ with respective BG regime curves and their total $F_{TBG}(T,n_s)= F_{DPBG}(T,n_s)+ F_{VPBG}(T,n_s)$. (b) for $n_s$=0.5-50.0x$10^{12}$ cm$^{-2}$.

It is to be noted that the BG regime studies in conventional in 2DEG [22],TMDs [23] and bilayer graphene (unscreened DP coupling) [18], with the parabolic dispersion of electron energy, show $n_s^{-3/2}$ dependence. But the $n_s^{-1/2}$ dependence of VP coupling in BLG, with the same parabolic dispersion for electrons, is due to the explicit dependence of its matrix element on electron energy (see Eq. (34) of Ref. [15]). Probably, this is the only matrix element which depends upon the initial state of the electron. The total $F_T(T,n_s)$ shows the $n_s^{-\delta(ns)}$ where $\delta(n_s)$ is a $n_s$ dependent positive exponent. In going from the $F_{DP}(T,n_s)$ dominant region to the $F_{VP}(T, n_s)$ dominant region, by increasing $n_s$, the $\delta(n_s)$ decreases from 3/2 to 1/2. This may be found from the curve for $F_T(T,n_s)$ in Fig. 2b.

### 3.2. Temperature dependence of $F(T,n_s)$

The temperature ($T$ = 0.5-50 K) dependence of $F_{DP}(T,n_s)$ and $F_{VP}(T,n_s)$ are shown in Fig. 3 for $n_s=1 \times 10^{12}$ cm$^{-2}$. This $T$ dependent behaviour has been explained in our earlier work [18]. The unscreened deformation potential theory of our calculations [18] are quantitatively compared with the experimental observations in Ref. [20]. With the choice of $D$ = 22eV an excellent agreement has been obtained. We see from Fig. 3 that for $n_s= 1 \times 10^{12}$ cm$^{-2}$, which is closer to the $n_s$ values in the samples of Ref. [20], the $F_{VP}(T,n_s)$ is nearly thirty times smaller than the $F_{DP}(T,n_s)$. The $F_T(T,n_s)$ is found to be dominated by $F_{DP}(T,n_s)$ in almost all over the temperature range considered. Hence signature of $F_{VP}(T, n_s)$ may not be noticed in the experimental observation of $F(T, n_s)$ in Ref. [20]. In MLG, the calculations of $F_{DP}(T,n_s)$ and $F_{VP}(T,n_s)$ as a function of temperature also show similar behaviour [10] in the clean limit. In this case, $F_{VP}(T, n_s)$ is shown to be about fifty times smaller than the unscreened $F_{DP}(T,n_s)$ for $D$= 20 eV. We suggest, to notice the significance of the $F_{VP}(T, n_s)$ measurements as a function of $T$ for larger $n_s$ (say $n_s > 1 \times 10^{13}$ cm$^{-2}$) need to be made.

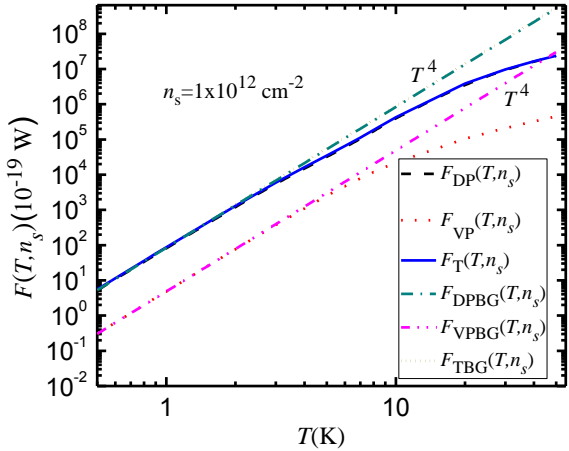

**Figure 3:** $F(T,n_s)$ vs $T$ for $n_s$ =1x10$^{12}$ cm$^{-12}$ due to DP and VP coupling and their total $F_T(T,n_s)= F_{DP}(T,n_s)+ F_{VP}(T,n_s)$. Respective BG regime curves with their total $F_{TBG}(T,n_s)= F_{DPBG}(T,n_s)+ F_{VPBG}(T,n_s)$ are also shown

In the BG regime $F_{VP}(T, n_s) \sim T^4$ which is same as that of $F_{DP}(T,n_s)$, and it is signature of 2D phonons. Same $T^4$ dependence is expected for both the $F_{VP}(T, n_s)$ and $F_{DP}(T,n_s)$, because $g_{DP}(q,k)$ and $g_{VP}(q,k)$, appearing in the respective matrix elements, turned out to be independent of $q$ in this region. The deviation from the $T^4$ law occurs for temperatures greater than about 3K in both the cases. However, the range of $T$ for the validity of power law is found to be marginally larger for $F_{VP}(T, n_s)$ than the $F_{DP}(T,n_s)$. This may be attributed to the behaviour of $g_{DP}(q,k)$ and $g_{VP}(q,k)$.

It is to be noted that, in $F_{VP}(T, n_s)$ the contribution comes from both LA and TA phonons. Since $v_l > v_t$, consequently, $F_{VP}(T, n_s)$ due to TA phonons is greater than that due to LA phonons. In the BG regime, $F_{VPTA}> F_{VPLA}$ by a factor of $(v_l/v_t)^3$, because $F_{VP}(T, n_s) \sim v_s^{-3}$.

Screening of the electron-phonon interaction affects only DP coupling [15]. To take account of screening, its matrix element has to be divided by the square of the long-wavelength static screening function $\varepsilon (q) = 1 + q_s/q$, where $q_s = q_{TF} = gme^2/\varepsilon \hbar^2$, is the 2D Thomas-Fermi wave vector [17]. In BG regime this enhances the power law for the temperature dependence and changes from the unscreened $T^4$ to $T^6$. Moreover, screening reduces $F_{DP}(T,n_s)$ by 2-3 orders of magnitude. This change due to screening of DP coupling will not account for the measured power loss [20] either by $T$ dependence or by magnitude. The contribution from the $F_{VP}(T,n_s)$ alone is much smaller than the measured value. Moreover, since the screening will not change the $n_s$ dependence, the $F_{VP}(T,n_s)$ fails to explain the observed $n_s^{-3/2}$ dependence.

In our equations (2) and (3), it is found that $F_{DP}(T,n_s) \sim m^2$ and $F_{VP}(T,n_s) \sim m^0$ (noting that $D_1 \sim m^{-1}$). Hence, any change in value of $m$ may cause significant change in $F_{DP}(T,n_s)$. Here we note that, in the literature, there is uncertainty in the value of $m$ and it ranges from 0.026 to 0.054 $m_0$ [15,16,24]. In BLG effective mass is given by $m=\gamma_1/2v_f^2$, where $\gamma_1$ is the interlayer coupling strength. Different groups of researchers use different values of $\gamma_1$ and $v_f$ leading to different $m$ values.

### 3.3. Relation between $F(T,n_s)$ and phonon limited mobility $\mu_p$ in BG regime

The low temperature acoustic phonon limited mobility $\mu_p$ is derived with the same basic assumptions as that in $F_i(T, n_s)$ with el-ap coupling as the common key element. $F_i(T, n_s)$ and $\mu_p$ are, respectively, due to the energy and momentum relaxation of the carriers. Hence, a relation between the two is expected. We obtain the acoustic phonon limited momentum relaxation time using the basic equation $(1/\tau_i)=\sum_{k'} (1-cos\theta_{kk'})W_{kk'}\{[1-f(E_k)]/[1-f(E_{k'})]\}$, where $W_{kk'}$ is the transition probability. In the BG regime, at Fermi energy $E_f$, the momentum relaxation times due to DP and VP coupling are, respectively, given by $(1/\tau_{DP}) = [D^2m(k_BT)^4!\zeta(4)]/[2\rho\pi^{5/2}\hbar^6v_s^5n_s^{3/2}]$ and $(1/\tau_{VP}) = [D_1^2m(k_BT)^4!\zeta(4)]/[\rho\pi^{3/2}\hbar^6v_s^5n_s^{1/2}]$. The acoustic phonon limited mobility $\mu_{pi}(T, n_s)$ is obtained using the relation $\mu_{pi}(T, n_s) = e\tau_i/m$, where $e$ is the electron charge. We see that the low-temperature equations for $F_i(T, n_s)$ and $\mu_{pi}(T, n_s)$ have many common factors. We obtain a simple relation of their product $F_i(T, n_s) \mu_{pi}(T, n_s) = \eta ev_s^2$ for both the DP and VP coupling, where $\eta$ is a numerical constant. In the present case it is found to be $\eta = 0.5$. We believe that this is an important relation connecting two measurable transport properties and the measurement of $F_i(T, n_s)$ will help to determine $\mu_{pi}(T, n_s)$ in the BG regime. Otherwise, it is difficult to measure $\mu_{pi}(T, n_s)$ due to contribution from the non-acoustic phonon mechanisms at low temperature.

We have also proved this relation between $F_i(T, n_s)$ and $\mu_{pi}(T, n_s)$ to be valid in three - and two-dimensional semiconductors, monolayer graphene and three-dimensional Dirac semimetals [25] with $0 < \eta < 1.5$. In Table 1, we have given the exponents of power laws for $T$ and $n_s$, along with the $v_s$, dependence, and $\eta$ values in different electron systems in BG regime. We note that the $n_s$ ($T$) dependent power law is determined by the dimensionality and dispersion of the electron (phonon) system. Interestingly, $\eta = 0.5$ found in BLG is same as that for 2DEG in MLG and monolayers of transition metal dichalcogenides [25]. This relation between $F(T, n_s)$ and $\mu_p(T, n_s)$ is analogous to Herring's law [26], which relates phonon-drag thermopower $S^g$ wth $\mu_p$, but with the difference that the present relation is independent of $T$, $n_s$ and other material parameters except $v_s$. Herring's law has been used to evaluate $\mu_p$ of electrons and composite fermions as a function of temperature by measuring the $S^g$ of 2DEG at zero magnetic field and composite fermions at high magnetic fields in GaAs/Ga$_{1-x}$Al$_x$As heterojunctions [27]. However, the present relation between $F(T, n_s)$ and $\mu_p(T, n_s)$ has an edge over the Herring's law in the systems in which $S^g$ is not observed ( for eg. in graphene with ~ few hundreds of nm size [32]) or not separable from the diffusion component of the thermopower. Moreover, the measurements of $F(T, n_s)$ are versatile and, in recent years, there are a large number of experimental studies being made in graphene ( for eg. see Ref [13]and references there in)].

Table 1. Power laws for $T$ and $n_s$, along with $v_s$ dependence, expressing $F_i(T, n_s)$ and $\mu_p(T, n_s) \sim T^\alpha$, $n_s^\delta$ and $v_s^\gamma$, where $\alpha$, $\delta$ and $\gamma$ are exponents. These $\alpha$, $\delta$ and $\gamma$ are constants in the Bloch-Grüneisen regime (otherwise they are generally function of $T$ and $n_s$) and are given for different electron systems and scattering mechanisms [25].

| Electron System | Power loss $F(T)$ | | | Acoustic phonon limited mobility $\mu_p$ | | | $\eta$ |
|---|---|---|---|---|---|---|---|
| | $\alpha$ | $\delta$ | $\gamma$ | $\alpha$ | $\delta$ | $\gamma$ | |
| 3D semiconductor: Deformation potential coupling | 5 | -1 | -4 [28]* | -5 | 1 | 6 [28] | 0.191 |
| 3D semiconductor: Piezoelectric coupling | 3 | -1 | -2 [28]* | -3 | 1 | 4 [28] | $1/\pi$ |
| 3D Dirac semimetal: Deformation potential coupling | 5 | -1/3 | -4 [29] | -5 | 1/3 | 6 [25] | 0.061 |
| Semiconductor HJs: Deformation potential coupling | 5 | -3/2 | -4 [22] | -5 | 3/2 | 6 [30] | 0.8 |
| Semiconductor HJs: Piezoelectric coupling | 3 | -3/2 | -2 [22] | -3 | 3/2 | 4 [30] | 1.332 For $v_t/v_l$= 0.59 [22,30] |
| Monolayer of transition metal dichalcogenides: Deformation potential coupling | 4 | -3/2 | -3 [23] | -4 | 3/2 | 5 [25] | 0.5 |
| Monolayer graphene: Deformation potential coupling | 4 | -1/2 | -3 [1] | -4 | 1/2 | 5 [31]** | 0.5 |
| Bilayer graphene: Deformation potential coupling | 4 | -3/2 | -3 | -4 | 3/2 | 5 | 0.5 |
| Bilayer graphene: Vector potential coupling | 4 | -1/2 | -3 | -4 | 1/2 | 5 | 0.5 |

\* In Eq.(8.27) of Ref.[28], in the denominator, the electron concentration $n_v$ seems to be missing.

\**In monolayer graphene, $\mu_p$ is obtained using $\tau_p(E_f)$ from Ref. [31].

**4. Conclusions**

We have given an analytical expression for the hot electron cooling power $F_{VP}(T, n_s)$ due to acoustic phonons through the VP coupling along with $F_{DP}(T, n_s)$ due to the DP coupling in a bilayer graphene at low temperature. Their relative contributions are discussed as a function of electron concentration $n_s$ and temperature $T$. Due to the different of $n_s$ dependence, the cross over between the two contributions is predicted with $F_{VP}(T, n_s)$ dominating at large $n_s$ ( for about >10$^{13}$ cm$^{-2}$). It is also found that both $F_{VP}(T, n_s)$ and $F_{DP}(T, n_s)$ have nearly the same temperature dependence. For the moderate $n_s$ (~10$^{12}$ cm$^{-2}$), the contribution from DP coupling is dominant and is in agreement with the experimental observations. A relation between $F(T,n_s)$ and phonon limited mobility $\mu_p$ is found in BG regime, which helps to determine $\mu_p$ from the measurements of $F(T,n_s)$. We emphasize that the $n_s$ dependent measurements of $F(T, n_s)$ are more important to identify the contribution due to VP coupling, significance of the screening and the Dirac phase of the electrons.